\documentclass[aps, prl,  twocolumn, preprintnumbers]{revtex4-1}

\let\counterwithout\relax

\pdfoutput=1
\usepackage{myoptions}

\usepackage{bbm}
\usepackage[caption=false]{subfig}

\usepackage{chngcntr}

\counterwithout{equation}{section}

\begin{document}

\title{The Quantum Null Energy Condition and Entanglement Entropy in Quenches}

\author{M\'ark Mezei}
\affiliation{
Simons Center for Geometry and Physics, SUNY, Stony Brook, NY 11794, USA}
\author{Julio Virrueta}
\affiliation{
C.N. Yang Institute for Theoretical Physics, SUNY, Stony Brook, NY 11794, USA}

\begin{abstract}

\noindent The Quantum Null Energy Condition (QNEC) relates energy to the second variation of entropy in relativistic quantum field theory. We use the QNEC inequality to bound entanglement entropy in quenches. At early times the entanglement entropy grows quadratically in time, and the QNEC provides an upper bound on the prefactor. We demonstrate that the bound is tight, by showing that it is saturated in certain quench protocols: boundary state quenches in conformal field theories in any dimensions. In higher than two dimensions  we compute entanglement entropy using AdS/CFT.  Our results are the first purely field theoretic applications of the QNEC.

\end{abstract}

\maketitle

\section{Introduction}

Positivity of energy is an important consistency requirement of relativistic quantum field theories (QFTs). QFT allows for localized negative energy, but various integrals of the energy momentum tensor are positive. The QNEC quantifies how negative the (null) energy can be at a spacetime point,  by  relating it to the second variation of entanglement entropy \cite{QNEC2,HoloQNEC,Wall:2017blw,QNEC1,Ceyhan:2018zfg}, and hence is of significant conceptual importance in QFT.  

In this letter, we turn the logic of QNEC around and use the null energy to bound the second time derivative of the entropy in an  out of equilibrium setup. We study the entanglement entropy of subregions in quantum quenches: we take a closed system in a homogeneous out of equilibrium (short range entangled) excited state, let it unitarily evolve and ask how quantities of interest change with time. The quench setup provides insight into the process of thermalization, hence it is of central importance in high energy \cite{Calabrese,Hartman}, condensed matter \cite{bardarson2012unbounded,serbyn2013universal,Kim:2013bc}, and quantum information theory \cite{Hayden:2007cs,Lashkari:2011yi}, and in cold atom experiments \cite{kaufman2016quantum}.

 In a generic strongly coupled many-body system in highly excited states, entanglement entropy is very hard to compute, hence it is valuable to constrain its dynamics by providing the tightest possible bound on it: this is what we set out to achieve.    If the state created by the quench has time reflection symmetry at $t=0$, the entropy starts growing quadratically in time; for  flat entangling surfaces $\Sig$ (corresponding to half space or strip subregions) we have
\es{Sexp}{
S(t)=S_0+ s_2 A_\Sig t^2+O(t^4)\,.
}
Applying the QNEC to this setup we find that 
\es{s2Bound}{
s_2&\leq {\pi c\ov \hbar} (e+p)\,,
}
where $e$ is the energy density and $p$ is the pressure.
We show that this bound is tight by constructing states in conformal field theory (CFT), where $s_2= {\pi c\ov \hbar} (e+p)$. The bound \eqref{s2Bound} is our main result, but we also
consider generalizations of it below. 

In recent years it has proven a fruitful direction to take the $G_N\to 0$ limit of conjectured quantum gravity bounds to obtain field theory inequalities \cite{Casini:2008cr,Bousso:2014sda,Bousso:2014uxa,FaulknerANEC,HartmanANEC}. These inequalities can then be independently proven with field theory methods, and they yield new knowledge about field theories and can in turn teach us about semiclassical quantum gravity. The QNEC is the latest such inequality, and the bounds derived in this paper are the first purely field theoretic applications of it. We anticipate that applying the QNEC in time dependent setups could teach us valuable lessons about out of equilibrium dynamics.

\section{Bounding entropy using the QNEC}
To state the QNEC, we have to introduce the setup and some notation. We work in $d$-dimensional Minkowski space with coordinates $x^\mu$, and set $\hbar=c=k_B=1$: these can be restored by dimensional analysis. Let us choose an entangling surface $\Sig$ defined by embedding functions $x^{\mu} = X^{\mu}(y)$, where $y^i$ are the $d-2$ coordinates parametrizing the surface. We choose  an orthogonal null vector field $k^\mu(y)$ on $\Sig$ and let  
 $\lam$ be an affine parameter along the geodesics generated by $k^\mu$.\footnote{The null vector field $k^\mu(y)$ has to be such that sufficiently many derivatives of the extrinsic curvature tensor vanish in the $k^\mu$ direction: $k_\mu \nabla_{i_1}\cdots \nabla_{i_n} K^\mu_{jk}=0$.} For infinitesimal $\lam$ the surfaces $\Sig(\lam)$ defined by $x^\mu=X^\mu(y)+k^\mu(y) \lam$ are deformations of the original entangling surface. 

We will use the nonlocal version of the QNEC, which is obtained by integrating the local version over the surface $\Sig$:
\begin{equation}
\label{intQNEC}
2\pi \int d^{d-2}y\ \sqrt{h_{\Sigma}}\, \left <T_{\mu\nu}\right> k^{\mu}k^{\nu}  \geq \frac{d^2}{(d\lambda)^2}S[{\Sigma}(\lam)]\,,
\end{equation}
where $T_{\mu\nu}$ is the stress tensor, $S[{\Sigma}(\lam)]$ is the entanglement entropy across $\Sig(\lam)$, and both sides are evaluated in a state $\ket{\psi}$ that we left implicit.  Note that the nonlocal version of the QNEC remains local in time.\footnote{We note that a local version of the QNEC refers to a contact term in the variation of \eqref{intQNEC}, and is always saturated in CFTs with a twist gap \cite{Leichenauer:2018obf,Balakrishnan:2019gxl}. Importantly, the saturation of the local version does not imply that the nonlocal version \eqref{intQNEC}  is also saturated. }

Our aim is to get information about entanglement entropy across entangling surfaces $\Sig(\lam)$ that lie on a constant time slice.  This requirement restricts the regions whose entanglement entropy we are able to constrain with the current argument to be a half space or a union of parallel strips (and spheres in CFTs discussed later).\footnote{The requirement that the entangling surfaces  $\Sig(\lam)$ belong to a constant time slice restricts the time component of $k^\mu$ to be a constant, and the surface orthogonality condition fixes $k^\mu=\text{const}\cdot(1,{\bf \hat n})\,,$ where ${\bf \hat n}$ is the unit normal vector to the surface. Any non-flat surface will have an extrinsic curvature in the normal vector direction, $k_\mu K^\mu_{jk}\neq0$ violating the condition discussed in \cite{Note1}.} 

 Let us first consider the case when the surface $\Sigma$ is a flat plane localized in  $x^0=t,\, x^{1}=x$ and extended along $x^{i+1}=y^{i},\, (i=1,\cdots d-2)$, introduce light cone coordinates $x^{\pm} \equiv x^0 \pm x^{1}$, and choose $k_\mu=\p_+$. Evaluating \eqref{intQNEC} gives
\begin{equation}
\label{intQNEC2}
2\pi \int d^{d-2}y \ \left <T_{++}(t,x,y^i)\right>  \geq \p_{++}S(t,x)\,,
\end{equation}
where $S(t,x)$ is the entropy across a plane located at $(t,x)$. We further specialize to homogeneous states, in which $\p_x S(t,x)=0$, and   since the stress tensor is a conserved current, its one point function is time independent, $\le<T_{\mu\nu}(x^\mu)\ri>=\text{const}$. We have $\left<T_{++}\right> = \frac{1}{4}\left(\left<T_{00}\right>+\left<T_{11}\right>\right) = \frac{1}{4}\left( e+p\right)$  in terms of the energy density $e$ and pressure $p$. Then \eqref{intQNEC2} becomes
\es{intQNEC3}{
2\pi A_\Sig (e+p)\geq {d^2 S(t)\ov dt^2}\,.
}
The bound is most interesting for a state that has time reflection symmetry at $t=0$. The symmetry implies that $\dot{S}(0)=0$, and we can expand the entropy at small $t$ as in \eqref{Sexp}. From \eqref{intQNEC3} 
 the coefficient $s_2$ is bounded as announced in \eqref{s2Bound}.
We can repeat the same argument for the union of parallel strips. We make the choice $k=\p_+$ on each disconnected component of $\Sig$, use the Ansatz \eqref{Sexp} and arrive at the same bound \eqref{s2Bound}.

For shapes different from half space, we do not know how to bound the $s_2$ from the QNEC. However, if the initial state at $t=0$ is short range entangled with correlation length $\xi$, then for shapes of curvature radius $R\gg\xi$, we expect the Ansatz \eqref{Sexp} to be valid with $s_2$ bounded by \eqref{s2Bound}.

At later times, the entropy is known to grow linearly \cite{Calabrese,Hartman,Liu1,Liu2}, 
\es{Linear}{
S(t)=S_0+v_E\,s_\text{eq}\, A_\Sig\, t+\cdots\,, \qquad t\gg t_\text{eq}\,,
}
where $s_\text{eq}$ is the equilibrium entropy density in the evolving state $\ket{\psi(t)}$,  $v_E$ is the entanglement velocity, while $t_\text{eq}$ is the local equilibration time. This linear growth does not contribute to the right hand side of \eqref{intQNEC3}, which hence is trivially satisfied. (We provide improvement on this state of affairs in $d=2$ CFTs below.) The linear growth can in turn be bounded using strong subadditivity of entanglement entropy \cite{Casini:2015zua} and the monotonicity of relative entropy \cite{Hartman:2015apr,Mezei:2016wfz}. It would be very interesting to combine these bounds with the QNEC. 

\section{Generalization for CFTs}

In $d=2$ CFT there exists a stronger version of the QNEC \cite{Wall:2011kb,HoloQNEC}:
\es{2dQNEC}{
\left<T_{\mu\nu}(x)\right>k^{\mu}k^{\nu}\geq \frac{d^2}{d\lambda^2} S(x) +{6\ov c}\le(\frac{d}{d\lambda} S(x)\ri)^2\,,
}
where $S(x)$ is the entropy of half space ending on $x$ and $c$ is the central charge of the CFT. The same steps as in the analysis of the half space case above lead to no improvement on the bound \eqref{s2Bound} on $s_2$, since the new term in \eqref{2dQNEC} is zero at $t=0$. At later times, however, \eqref{2dQNEC} provides an important improvement over the bound \eqref{intQNEC3}, as will be shown when we analyze concrete quenches.

If the  system in question is a higher dimensional CFT, another version of the QNEC also exists that applies to spherical regions of radius $R$ \cite{HoloQNEC}:
\es{ConfQNEC}{
&2\pi R^{d-2}\int d\Omega \left<T_{\mu\nu}(\Omega)\right> k^{\mu}k^{\nu}\\
&\geq \frac{d^2}{d\lambda^2}\Delta S[\Sig(\lam)] - \frac{2}{R}\frac{d}{d\lambda}\Delta S[\Sig(\lam)] \,,
}
where $\Delta S[\Sig(\lam)] $ is the vacuum subtracted entanglement entropy of the sphere $\Sig(\lam)$, and   the null vector is taken to be in the inward radial direction, $k^\mu=(1,-{\bf \hat r})$, hence ${d\ov d \lam}=\p_t-\p_R$. In a time reflection symmetric state, we adapt \eqref{Sexp} to the sphere setup 
\es{Sexp2}{
\Delta S(t,R)=\Delta S_0(R)+ A_R\, s_2(R)\, t^2+O(t^4)\,,
}
where $A_R$ is the surface area of sphere. Because the entropy of the vacuum state does not change with time, we ommited the $\Delta$ from $s_2(R)$.
Since a state generically has an intrinsic correlation length scale, $\Delta S_0(R)$ and $ s_2(R)$ can have an arbitrarily complicated $R$ dependence.
Plugging this into \eqref{ConfQNEC} we obtain:
\es{s2BoundSphere}{
 s_2(R)+{1\ov A_R}\le({1\ov 2}\Delta S''_0(R)+{1\ov R}\Delta S'_0(R)\ri)\leq \pi (e+p) \,,
}
where $p=e/(d-1)$ due to the CFT equation of state.
 This bound is not as clean as in the half space case, but it may prove to be useful, if we know the entanglement structure of the initial state. If we assume that the initial state was a scale invariant state distinct from the vacuum, then $\Delta S_0'(R)=0$, and we get $s_2\leq \pi (e+p)$ that takes the same form as \eqref{s2Bound}.\footnote{Neither $\Delta S_0$ nor $s_2$ can depend on $R$ in such states.} We construct such a special excited state below.
 
  Another case when we can simplify \eqref{s2BoundSphere} is the $R\to\infty$ limit. Since no state is expected to have super-volume law entanglement, we have at most $S_0(R)\sim R^{d-1}$, which makes the terms coming from  $S_0(R)$ subleading to $s_2(R)$, and we again recover the formula $ s_2(\infty)\leq \pi (e+p)$. This result is in line with the discussion of large arbitrary shapes above.

\section{Quenches saturating the bound}

We have obtained a bound on $s_2$. A bound is the most interesting if it is tight, i.e.\@ there exists a setup when it is saturated. In such a favorable case we know we cannot improve the bound any further without making extra assumptions. We now show that the bound on $s_2$ \eqref{s2Bound} is tight. 

Let us consider a conformal field theory (CFT), and prepare the following out of equilibrium short-range entangled time reflection symmetric initial state \cite{Calabrese}:
\es{InitialState}{
\ket{\psi_0}= e^{-\beta/4\, H}\, \ket{B}\,,
}
where  $\ket{B}$ is a conformal boundary state and $\beta$ is the effective inverse temperature of the state, $\bra{\psi_0}H\ket{\psi_0}/\bra{\psi_0}\psi_0\rangle=E(\beta)$ \cite{CCreview}.\footnote{That the expectation value is $E(\beta)$ is only known to be true in $d=2$ CFT and higher dimensional holographic theories for the boundary states that we analyze: the two examples that we will be considering.} We time evolve this initial state according to $\ket{\psi(t)}=e^{-iHt}\ket{\psi_0}$ and want to understand the entanglement entropy of half space  in this state.

In $d=2$  Calabrese and Cardy determined this entanglement entropy by CFT techniques \cite{Calabrese}:
\es{2dCFT}{
S(t)&={c\ov 6}\log\le[{\beta\ov 2\pi \ep}\cosh \le(2\pi t\ov \beta\ri)\ri]\\
&={c\ov 6}\log\le(\beta\ov 2\pi \ep\ri)+{c \,\pi^2\ov 3\beta^2}\, t^2+O(t^4)\,,
}
 where we denoted the short distance cutoff by $\ep$, and  in the second line we expanded their answer for early time and defined $S_0={c\ov 6}\log\le(\beta\ov 2\pi \ep\ri)$. Using the equation of state $e=p={c\,\pi \ov 6\beta^2}$, we read off that 
\es{s22d}{
s_2= \pi (e+p)={c \,\pi^2\ov 3\beta^2}\,,
}
saturating the bound \eqref{s2Bound}. In fact, remarkably $S(t)$ saturates for all times the stronger version of QNEC \eqref{2dQNEC} valid in $d=2$ CFTs,
\es{Alltimes}{
\pi (e+p)=\ddot{S}(t) +{6\ov c}\, \dot{S}(t)^2\,.
}
This means that the stronger QNEC bound \eqref{2dQNEC} in $d=2$ CFT is tight for all times.\footnote{Returning to \eqref{Linear}, for this quench we have $v_E=1,\, s_\text{eq}={c\pi\ov 3\beta}$,  and the linear growth only contributes to the new second term in \eqref{Alltimes}.} 
The equality is only true for half space, but is expected to be a good approximation for intervals of width $L$ for $t< L/2$. 

\begin{center}
\begin{figure}[!h]
\includegraphics[scale=0.7]{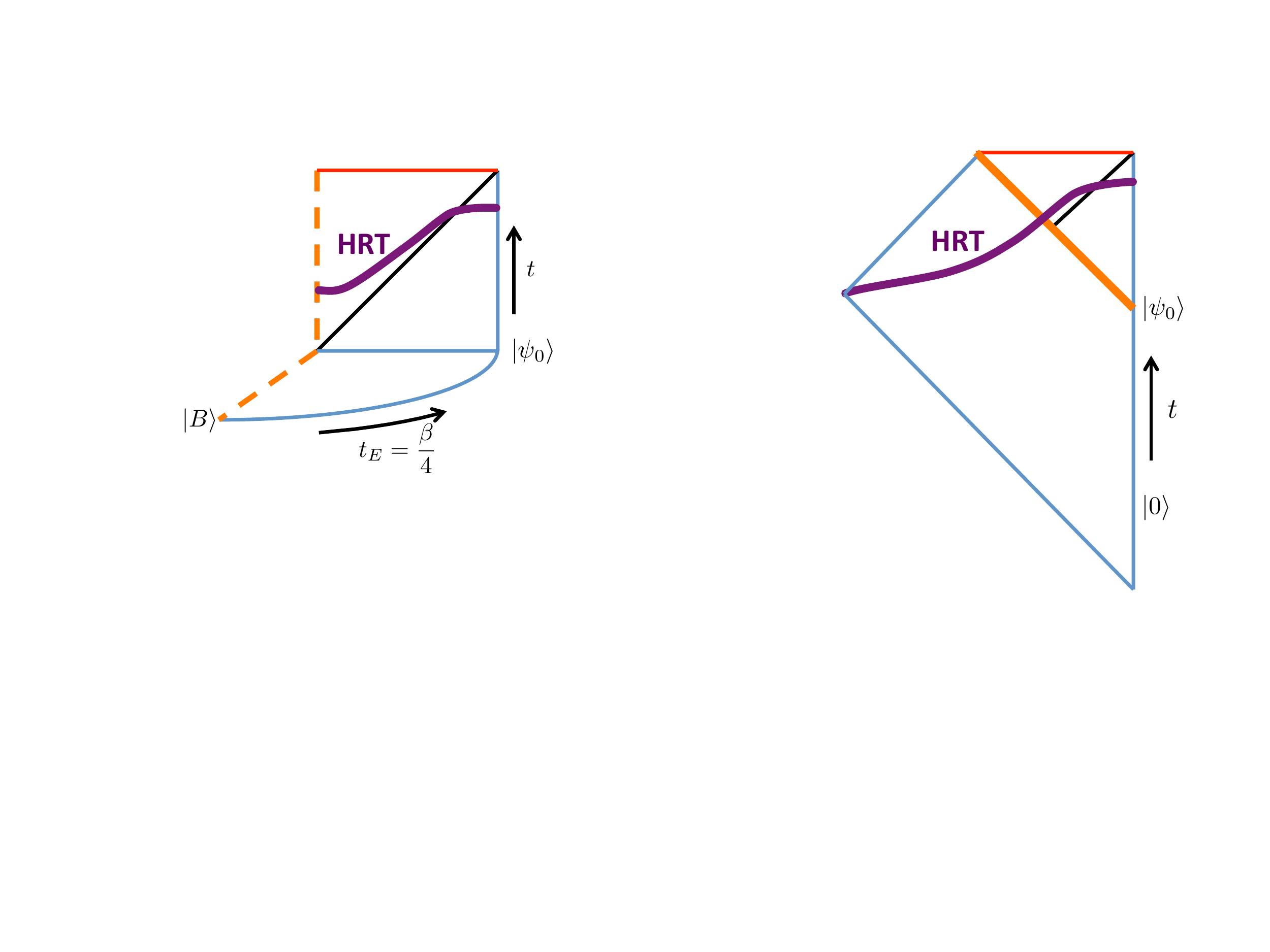}
\caption{Penrose diagram of the spacetime dual to the boundary state quench state $\ket{\psi(t)}$. Lorentzian time flows upwards, while Euclidean time $t_E$ is drawn perpendicular to the page. The geometry is  that of an eternal black brane cut in half by an end of the world brane. The horizon is a diagonal black line, the singularity is a red line, the $t=0$ hypersurface and the AdS boundary are drawn by blue, while the end of the world brane is a dashed orange line. The HRT surface computing the entropy of half space is a purple curve: its nontrivial motion is in the $(t,z)$ plane, while it stays a flat plane in the ${\bf x}$ directions that are suppressed on the figure.
\label{fig:Penrose}}
\end{figure}
\end{center}

In higher dimensional CFTs conformal symmetry is less powerful, and we do not know how to obtain $S(t)$ in general. For CFTs that have a holographic dual description, however, the HRT prescription \cite{Ryu:2006ef,Ryu:2006bv,Hubeny:2007xt} provides an elegant method for computing the entanglement entropy as the area of an extremal surface in the geometry  dual to the state $\ket{\psi(t)}$, which is the eternal black brane cut in half by an end of the world brane \cite{Hartman}:\footnote{Here we chose the brane to be tensionless. The brane tension is set by the boundary state $\ket{B}$, see \cite{Kourkoulou:2017zaj,Almheiri:2018ijj,Cooper:2018cmb} for spacetimes with branes that have nonzero tension.  }
\begin{equation}
\label{metric}
ds^2 = \frac{1}{z^2}\left(-{a(z)}dt^2 + d{\bf x}^2+{dz^2\ov a(z)b(z)^2}\right)\,,
\end{equation}
where $t,{\bf x}$ are the field theory and $z$ the bulk radial coordinate, $a(z),\,b(z)$ are functions constrained by the AdS boundary conditions at $z=0$ and the null energy condition (NEC). Without loss of generality we put the horizon at $z_h=1$, where $a(1)=0$. The most natural geometry corresponds to the AdS Schwarzschild black brane, $a(z)=1-z^d,\, b(z)=1$: this geometry is expected to model a family of boundary states $\ket{B}$. See Fig.~\ref{fig:Penrose} for the Penrose diagram and the sketch of the HRT surface that ends on the end of the world brane.

The entanglement entropy is computed by the parametric function $(t(z_*),S(z_*))$, where $z_*$ is the location, where the HRT surface ends on the end of the world brane, given by the integrals
\es{tSInts}{
t (z_*) &= \frac{i \pi}{a'(1)b(1)} - \int_0^{z_*} \frac{dz}{a(z)b(z)\sqrt{1-\frac{a(z)}{a(z_*)}\left(\frac{z_*}{z}\right)^{2(d-1)}}}\,,\\
 S(z_*) &= \frac{A_{\Sigma}}{4G_N}\int_\ep^{z_*} \frac{dz}{z^{d-1}}\frac{1}{b(z)\sqrt{a(z) - a(z_*)\left(\frac{z}{z_*}\right)^{2(d-1)}}}\,,
}
where $G_N$ is Newton's constant in the AdS gravity theory. We provide the derivation of these formulas as well as their detailed analysis in the Supplemental Material (SM). The integrals can be evaluated at early time corresponding to $z_*\sim 1$:
\es{tSInts2}{
t (z_*) &=C_1 \sqrt{z_*-1}+O(z_*-1)\,,\\
 \hat S(z_*) &= C_2 \, A_{\Sigma}\,\le(z_*-1\ri)+O\le((z_*-1)^2\ri)\,,
}
where $ \hat S(z_*(t))\equiv S(t)-S(0)$ and $C_i$ are constants that are functionals of  $a(z),\,b(z)$ and are given in the SM. From \eqref{tSInts2} we obtain
\es{tSInts3}{
 \hat S(t) &= {C_2\ov C_1^2}  \, A_{\Sigma}\, t^2+O\le(t^4\ri)\ \implies \ s_2^\text{(holo)}={C_2\ov C_1^2}\,.
}

The combination $\pi (e+p)$ can be read off from the near boundary behavior of the geometry \eqref{metric} according to the holographic dictionary \cite{Balasubramanian}:
\es{epHolo}{
a(z)&=1 - a_d z^d+\cdots\,,\\
\pi (e+p)&={d a_d\ov 16 G_N}\,,
}
and in the SM we show that the NEC implies
\es{HoloIneq}{
s_2^\text{(holo)}\leq \pi (e+p)\,,
}
with equality only for the AdS Schwarzschild black brane. We note that in $d=2$ the holographic formula recovers the field field theory result \eqref{s22d}.

Our result achieves two goals at once. First, it provides an alternative, holographic derivation of the bound \eqref{s2Bound} that was obtained from the QNEC. Second, it shows that the bound is tight: it is saturated in holographic CFTs in the boundary state quench \eqref{InitialState} (for $\ket{B}$ such that we get the Schwartzschild geometry).

We remark that there is a curious example of colliding shockwaves in holography, where the saturation of the nonlocal version of the QNEC was observed numerically \cite{Ecker:2017jdw}. In holography in $d=2$ many states obeying the analog of  \eqref{Alltimes} were found in \cite{Khandker:2018xls,Ecker:2019ocp}. Note, however, that \eqref{Alltimes} is true in any $d=2$ CFT not just in those with semiclassical holographic duals.

\section{Another special quench protocol}

\begin{center}
\begin{figure}[!h]
\includegraphics[scale=0.6]{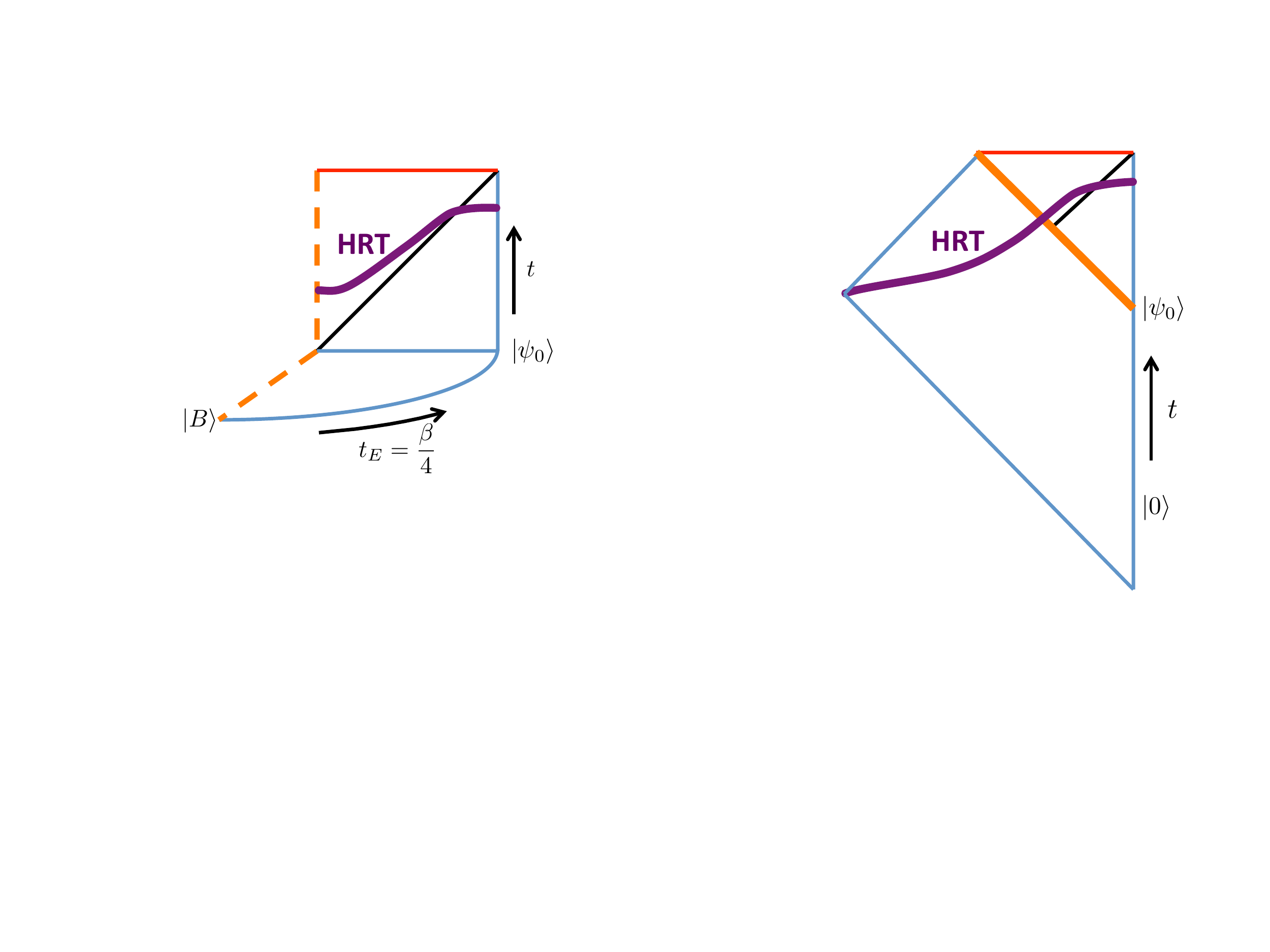}
\caption{Penrose diagram of a Vaidya spacetime. The infalling matter shell is a diagonal orange line, and we glue a pure AdS spacetime (dual to $\ket{0}$) to a black hole spacetime across it.
\label{fig:Penrose2}}
\end{figure}
\end{center}

In gravity it is natural to consider the Vaidya spacetime that describes a very thin collapsing null dust shell that forms a black brane. The spacetime is obtained by gluing pure AdS spacetime to a black brane spacetime across a null hypersurface, see Fig.~\ref{fig:Penrose2} and the SM. 

The dual CFT state is obtained by acting on the vacuum with a product of operators $\sO$ inserted in small Euclidean time $t=ia$ and distributed on a lattice whose lattice constant $a$ we take to be small \cite{Anous:2016kss,Anous:2017tza}:
\es{VaidyaQFT}{
\ket{\psi_0}=\le(\prod_{{\bf n}\in \Z^{d-1} } \sO(ia, a {\bf n})\ri) \ket{\psi_0}\,.
}
This state has energy density $e\propto {\De_\sO\ov  a^{d}}$, and has an approximately scale invariant entanglement structure for regions with $R\gg a$.\footnote{To get a more precise match with the Vaidya geometry, in the large $c$ limit we take the limits $c\to\infty,\,{\De_\sO/ c} =\text{fixed},\,{e/ c} =\text{fixed}$ first, followed by  $a\to0,\, {\De_\sO/ c} \to 0,\,  {\De_\sO/ (c\, a^{d})}\propto {e/ c}=\text{fixed}$.} This is a special scale invariant excited state we were looking for to make the bound for spheres \eqref{s2BoundSphere} easier to apply.

Using holographic duality,   following \cite{Liu1,Liu2} in the SM we compute the early time growth of entanglement entropy across arbitrary entangling surfaces $\Sig$. The result is
\es{EarlyVaidya}{
S[\Sig(t)]=S_0[\Sig]+ s_2^\text{(holo)} A_\Sig t^2+O(t^4)\,,
}
where $s_2^\text{(holo)}$ is independent of the shape, and $S_0[\Sig]$ depends on the shape but not on the overall size. Both our QFT bound for half space \eqref{s2Bound} and our CFT bound for spheres \eqref{s2BoundSphere} can be used to bound $s_2^\text{(holo)}\leq \pi (e+p)$.

It turns out that the bound is too loose for these states. Using the NEC, in the SM we give a holographic proof that in the state \eqref{VaidyaQFT} dual to Vaidya spacetimes obeys
\es{s2Vaidya2}{
s_2^\text{(holo)}\leq {\pi \, (e+p)\ov d}\,,
}
which is a factor of $d$ stronger than the QNEC bounds \eqref{s2Bound} and \eqref{s2BoundSphere}.\footnote{In the main text, we evolve the state $\ket{\psi_0}$ \eqref{VaidyaQFT} with a CFT Hamiltonian, hence $p=e/(d-1)$ by the equation of state, but in the SM we consider a more general setup, where we evolve with a possibly massive QFT Hamiltonian.}  For $d=2$ the factor of 2 improvement over \eqref{s2Bound} in \eqref{s2Vaidya2} was noticed in \cite{Ecker:2019ocp}. It would be very interesting to find an alternative proof method in field theory to demonstrate \eqref{s2Vaidya2} for the special class of states considered here.\footnote{There is a proof strategy \cite{casini_unpub} based on the positivity of relative entropy which in its current incarnation can prove $s_2\leq \pi (e+p)$ for states of the type \eqref{VaidyaQFT}, but that does not apply to the boundary state quenches \eqref{InitialState} that were the most interesting for this paper. This alternative strategy might perhaps be improved to show \eqref{s2Vaidya2}.}

\section{Conclusions and outlook}

In this paper, we used the QNEC, a novel insight into the relation between energy and entropy in QFT to bound  the entanglement entropy in quenches. The bounds are most interesting at early times (see however \eqref{Alltimes} in $d=2$ CFT), where we have shown that the conformal boundary quench protocol saturates the bound \eqref{s2Bound}. 

The bound on $s_2$ applies to any experimental or theoretical setup, where the Hamiltonian and the quench state are close to the continuum limit that is described by a relativistic QFT, hence it can provide a valuable consistency check of these results.\footnote{The value of $c$ should be taken to be the emergent speed of light.} It also provides the maximum value for $s_2$, to which we can compare the measured or computed value.

In the future it would be interesting to extend the bounds to other shapes by perhaps including more information about the entanglement structure of the initial state. Another option is to exploit conformal symmetry in CFTs efficiently, as in \cite{HoloQNEC}. There has been interesting developments on entanglement dynamics in the ``hydrodynamic'' regime $R,t\gg t_\text{eq}$ \cite{Hartman,Liu1,Liu2,Jonay:2018yei,Mezei:2018jco,Zhou:2018myl,Rakovszky:2019oht,Kudler-Flam:2019wtv,Wang:2019ued}, whereas our work offers insights at earlier times. It would be very interesting to combine these approaches to obtain a better understanding of entanglement growth in quenches. The first step in this direction can be taken by combining the bound \eqref{s2Bound} with the linear regime of entropy growth \eqref{Linear}. We obtain a lower estimate on the local equilibration time:
\es{teq}{
t_\text{eq}\approx {c\, s_\text{eq}\ov k_B \, s_2} \gtrsim {\hbar\, s_\text{eq}\ov  k_B\, e}\approx {\hbar\ov  k_B T}\,,
}
which interestingly reproduces the well known Planckian lower bound on $t_\text{eq}$. We expect new insights about out of equilibrium dynamics to emerge from the interplay of these disparate tools.\\*

{\it Acknowledgments:} We thank John Cardy, Horacio Casini, and especially Tom Faulkner for useful discussions.  MM is supported by the Simons Center for Geometry and Physics. JV is supported by NSF award PHY-1620628.

\bibliography{QNECRefs}

\clearpage

\onecolumngrid
\begin{center}
\vspace{0.5cm}
\textbf{\large Supplemental Material for ``The Quantum Null Energy Condition and Entanglement Entropy in Quenches''}\\[5pt]
\end{center}

\twocolumngrid


%
%


%
%

\appendix
\section{Holographic entanglement entropy for the boundary state quench}

We present the derivation of  equation \eqref{tSInts}, which gives the parametric function $(t(z_*),S(z_*))$. As explained in the main text, the boundary state quench is described by a black brane geometry cut in half  by an end of the world brane. The geometry is given in \eqref{metric} and is written below in infalling coordinates:
\begin{equation}
\begin{aligned}
ds^2 &= \frac{1}{z^2}\left(-a(z)dv^2 - \frac{2}{b(z)}dvdz+d\vec{x}^2\right)\,,\\
v&\equiv t - \int_0^z \frac{dz'}{a(z')b(z')}\,.
\end{aligned}
\end{equation}
The time coordinate is extended behind the horizon as $t = t_I + \frac{i\pi}{a'(1)b(1)}$. Since the end of the world brane is located at the fixed plane of time reflection symmetry $t_I\to-t_I$, its position in $v$ is given by:
\begin{equation}
v_{\text{brane}}(z) = \frac{i\pi}{a'(1)b(1)} - \int_0^z \frac{dz'}{a(z')b(z')}\,.
\end{equation}

The entanglement entropy is computed as the area of an extremal codimension-two surface that ends on the end of the world brane behind the horizon.  When the subregion of interest is a half space, the HRT surface only moves nontrivially in the $(t,z)$ plane. Therefore its embedding is given by $z=z(v)$. The area functional was computed  in  \cite{Hartman,Liu1,Liu2,Mezei:2016zxg}:
\begin{equation}
S(t) = \frac{A_{\Sigma}}{4G_N}\int_{v_{\text{brane}}(z_*)}^{t} dv\ \frac{\sqrt{Q}}{z(v)^{d-1}}\,, \quad Q \equiv a(z) + \frac{2z'(v)}{b(z)}\,, 
\end{equation}
where $z_*$ is the point where the HRT surface ends on the end of the world brane.
Since the functional is independent of $v$, we have a conserved energy:
\begin{equation}
E = \frac{1}{z^{d-1}\sqrt{Q}}\left(a(z) + \frac{z'(v)}{b(z)}\right)\,,
\end{equation}
which allows us to write the boundary time and the area as
\es{tSIntsApp}{
t (z_*) &= \frac{i \pi}{a'(1)b(1)} - \int_0^{z_*} \frac{dz}{a(z)b(z)\sqrt{1-\frac{a(z)}{a(z_*)}\left(\frac{z_*}{z}\right)^{2(d-1)}}}\,,\\
 S(z_*) &= \frac{A_{\Sigma}}{4G_N}\int_\ep^{z_*} \frac{dz}{z^{d-1}}\,\frac{1}{b(z)\sqrt{a(z) - a(z_*)\left(\frac{z}{z_*}\right)^{2(d-1)}}}\,.
}
This is \eqref{tSInts}, as announced in the main text.

\section{Holographic entanglement entropy at early times}

The next step is to study the integrals \eqref{tSIntsApp} in the early time regime, which corresponds to the limit $z_*\rightarrow 1^+$. It is important to notice that the integrals in both expressions are singular at $z=1$ and must be regularized. To do this we divide the contour of integration into three pieces: a section $z\in[0,1-\delta]$  outside the horizon, an additional section $z=1-\delta\,  e^{i\theta}$, $\theta\in[0,\pi]$ around it in the complex plane, and a final section $z\in[1+\delta,z_*]$ inside the horizon.

We take  the following order of limits: $0<\delta\ll (z_*-1)\ll 1$. For the integral $t(z_*)$, the piece from the middle section of the contour cancels the first term $\frac{i \pi}{a'(1)b(1)} $, while for $S(z_*)$ this middle section contributes only at $O(\delta)$. We expand the remaining integrals for small $\delta$. In both cases, the integral inside the horizon contributes only a divergent piece, which cancels exactly the divergence of the integral outside the horizon. We obtain
\bwt
\begin{equation}
\begin{aligned}
t_b (z_*) &=  - \sqrt{-a'(1)(z_* -1)}\left(\int_0^{1-\delta} dz \ \frac{z^{d-1}}{b(z)a(z)^{3/2}}  - \frac{2}{b(1)(-a'(1))^{3/2}\sqrt{\delta}}\right)+O(z_*-1)\,,\\
\hat{S}(z_*) &= \frac{a'(1)}{8G_N}A_{\Sigma}(z_*-1)\left(\int_0^{1-\delta} dz \ \frac{z^{d-1}}{b(z)a(z)^{3/2}}- \frac{2}{b(1)(-a'(1))^{3/2}\sqrt{\delta}}\right)+O\le((z_*-1)^2\ri)\,,
\end{aligned}
\end{equation}
which is the expansion given in the main text in \eqref{tSInts2}. The constants $C_i$ appearing in  \eqref{tSInts2} have the explicit expression:
\begin{equation}
\begin{aligned}
C_1 &=  - \sqrt{-a'(1)}\left(\int_0^{1-\delta} dz\ \frac{z^{d-1}}{b(z)a(z)^{3/2}}- \frac{2}{b(1)(-a'(1))^{3/2}\sqrt{\delta}}\right)\,,\\
C_2 &= \frac{a'(1)}{8G_N}\left(\int_0^{1-\delta} dz \ \frac{z^{d-1}}{b(z)a(z)^{3/2}} -  \frac{2}{b(1)(-a'(1))^{3/2}\sqrt{\delta}}\right)\,.
\end{aligned}
\end{equation}
We conclude from \eqref{tSInts3} that
\es{s2holoFinal}{
s^{\text{(holo)}}_2 &= \frac{C_2}{C_1^2} =  \frac{1}{8G_N}\left(-\int_0^{1-\delta} dz\ \frac{z^{d-1}}{b(z)a(z)^{3/2}}+ \frac{2}{b(1)(-a'(1))^{3/2}\sqrt{\delta}}\right)^{-1}\,.
}
\ewt

\section{Quadratic growth bound from the NEC}

First, we consider the NEC for the bulk matter computed from the geometry, $T^{\text{bulk}}_{\mu\nu}\ell^{\mu}\ell^{\nu}\geq0$, where $T^{\text{bulk}}_{\mu\nu}$ is the gravitational stress tensor, and $\ell^{\mu}$ is an arbitrary null vector in the bulk. Outside the horizon, $0\leq z \leq1$, the NEC translates into \cite{Mezei:2016zxg}:
\es{NEC}{
\frac{d}{dz}\left(\frac{b(z)a'(z)}{z^{d-1}}\right) &\geq 0\,,\\
b'(z) &\geq 0\,,
}
which can be enforced by considering it as a system of ordinary differential equations \cite{Mezei:2016zxg}:
\begin{equation}
\label{NECBH}
\begin{aligned}
\frac{d}{dz}\left(\frac{b(z)a'(z)}{z^{d-1}}\right) &= s(z)\,,\\
b'(z) &= t(z)\,,
\end{aligned}
\end{equation}
where the sources $s(z)$ and $t(z)$ positive due to the NEC \eqref{NEC}.

We can then solve these equations with the boundary conditions $a(0)=b(0)=1$ that impose AdS asymptotics, and $a(1)=0$ that fixes the location of the horizon  \cite{Mezei:2016zxg}:
\es{metricSol}{
b(z) &= 1 + \int_0^z dz' \ t(z') \,,\\
B(z) &\equiv \int_0^z dz' \ \frac{z'^{d-1}}{b(z')}\,,\\
a(z) &= \left(1 - \frac{B(z)}{B(1)}\right)\left(1 - \int_0^1 dz' \ B(z')s(z')\right) \\
&- \int_z^1 dz'\ \left(B(z)- B(z')\right)s(z')\,.
}
Since $B(z)$ is a monotonically increasing function of $z$, the last term in $a(z)$ is always positive for $0\leq z<1$. Furthermore, from the requirement that $z_h=1$ is the outer horizon of the black brane (and hence $a(z)>0$ in the region $0\leq z<1$), it follows that $a'(1)<0$, which implies $\int_0^1 dz' \ B(z') s(z') <1$. We will use these ingredients to bound $s^{\text{(holo)}}_2$ given in \eqref{s2holoFinal}.

Second, we  apply the holographic dictionary to obtain the stress tensor of the boundary theory \cite{Balasubramanian}:
\es{Tholo}{
T_{\mu\nu} &= \frac{1}{8\pi G_N} \lim_{z\rightarrow 0} \frac{1}{z^{d-2 }}\left[K_{\mu\nu} - \gamma_{\mu\nu} K \ri.\\
&\le.- (d-1)\gamma_{\mu\nu}+\text{(extra counter terms)}\right]\,,
}
where $\gamma_{\mu\nu}$ is the boundary metric, $K_{\mu\nu}$ is the extrinsic curvature tensor, $K=K^\mu_{\mu}$, and the terms in the second line are counter terms. We have written out the first universal counter term, and left the subleading ones implicit; holographic renormalization is the technology developed to deal with them \cite{Skenderis:2002wp}. We are interested in Poincare invariant field theories on flat space since the QNEC only applies to these. Therefore $\gamma_{\mu\nu}\propto\eta_{\mu\nu}$ and  all the counter terms are required by symmetry to be proportional to $\eta_{\mu\nu}$. To be specific, in case the CFT is deformed by a relevant operator $\sO$, the dual bulk scalar field profile would  contribute to the extra counter terms in \eqref{Tholo}. One can explicitly check that these extra counter terms are proportional to $\eta_{\mu\nu}$. The near boundary expansion of the metric functions was sketched in \eqref{epHolo}, which we repeat here.
\es{epHoloApp}{
a(z)&=1 - a_d z^d+\cdots\,,\\
b(z)&=1+ \# z^{p}+\cdots \,,
}
where the $z^{p}$ term is induced by the scalar field profile, with $p<d$. This term then contributes a divergent $z^{p-d}$ term to the first line of \eqref{Tholo} and is cancelled by the extra counter terms. All these details are irrelevant for our case, where the null projection of the stress tensor, $T_{\mu\nu}k^{\mu}k^{\nu}$, cancels all terms proportional to $\eta_{\mu\nu}$. We  conclude
\es{epHoloApp2}{
\pi (e+p)&={d a_d\ov 16 G_N}\\
&= \frac{1}{16 G_N \,B(1)}\left(1+ \int_0^1dz'\left(B(1) - B(z')\right)s(z')\right)\,,
}
where in the second line we used \eqref{metricSol}.

Finally, we put the pieces together. We use the expressions \eqref{metricSol} to evaluate $s^{\text{(holo)}}_2$ from \eqref{s2holoFinal}:
\bwt
\begin{equation}
\begin{aligned}
 \frac{1}{s^{\text{(holo)}}_2} \geq 8G_N \left(\frac{2B'(1)}{(-a'(1))^{3/2}\sqrt{\delta}} - \int_0^{1-\delta}dz\ \left(\frac{B'(z)}{\left(1-\frac{B(z)}{B(1)}\right)^{3/2}\left(1-\int_0^1dz' B(z') s(z')\right)^{3/2}}\right) \right)\,,
\end{aligned}
\end{equation}
\ewt
where in the integral we dropped the last term from the expression of $a(z)$ since it is always positive on the domain on integration as discussed below \eqref{metricSol}. We also used $B'(z) = \frac{z^{d-1}}{b(z)}$ to remove all explicit dependence on $b(z)$. The remaining integral can be evaluated exactly as:
\begin{equation}
\frac{1}{s^{\text{(holo)}}_2} \geq  \frac{16G_N B(1)}{\left(1 - \int_0^1 dz' B(z') s(z') \right)^{3/2}}\,.
\end{equation}
Since the denominator is always smaller than one (but cannot be zero as was explained below \eqref{metricSol}), we get the inequality:
\begin{equation}
s^{\text{(holo)}}_2 \leq \frac{1}{16G_N B(1)}\,.
\end{equation}
Comparing with \eqref{epHoloApp2}, and noticing that in \eqref{epHoloApp2} the term in parentheses is always larger than one, we obtain the bound:
\begin{equation}
s^{\text{(holo)}}_2 \leq \pi (e+p)\,,
\end{equation}
which is saturated if and only if $s(z)=t(z) =0$. This corresponds to the AdS Schwarzschild black brane geometry.

\section{Quadratic growth bound for the Vaidya spacetime}

In the SM we work with a more general class of Vaidya spacetimes than in the main text: we allow the spacetime before the infalling null shell to be the vacuum of a scale non-invariant holographic theory. In the Vaidya spacetime, the metric components are a function of both $z$ and the infalling coordinate $v$. The $v$ dependence comes from the modeling of the infalling dust. For $v>0$, the metric takes the form of the black brane geometry:
\begin{equation}
ds^2 = \frac{1}{z^2}\left(-a(z)dv^2 - \frac{2}{b(z)}dzdv + d{\bf x}^2\right)\,,
\end{equation}
while for $v<0$ we have the most general (boundary) Poincare invariant spacetime dual to a renormalization group flow:
\es{DomainWall}{
ds^2 = \frac{1}{z^2}\left(-dv^2 - \frac{2}{f(z)}dzdv +  d{\bf x}^2\right)\,,
}
where the function $f(z)$ obeys a null energy condition $f'(z)\geq 0$ along with the boundary condition $f(0)=1$. These two geometries are glued together along the null hypersurface $v=0$ and give the $v$-dependent metric:
\begin{equation}
ds^2 = \frac{1}{z^2}\left(-a(z,v)dv^2 - \frac{2}{b(z,v)}dzdv + d{\bf x}^2\right)\,,
\end{equation}
where $a(z,v)=1-(a(z)-1)\theta(v)$ and $b(z) = f(z)-(b(z)-1)\theta(v)$. The transition between the two is produced by a matter stress tensor localized on the null hypersurface, corresponding to null dust with surface energy density and pressure given by:
\es{NECShell}{
e_\text{dust}(z) &= \frac{z}{8\pi}\left(f(z)^2 - b(z)^2 a(z)\right)\,,\\
p_\text{dust}(z)&=\frac{1}{8\pi}\left(\frac{b'(z)}{b(z)}- \frac{f'(z)}{f(z)}\right)\,.
}
Demanding that this localized stress tensor obeys the NEC implies that these two quantities are always non-negative. For the   most often considered Vaidya setup, given by $b(z)=f(z)=1,\, a(z)=1-z^d $, the second condition is trivial, while the first condition is the familiar positivity constraint on the black brane mass.

We note that from the gravitational perspective, we are allowed to glue together the spacetime \eqref{DomainWall} with a black brane with $a(z)$ that in its small $z$ expansion  contains lower powers than $z^d$. As discussed around  \eqref{epHoloApp} such  terms are disallowed by Poincare invariance of the dual field theory. If we allowed such terms, they would lead to divergent change in energy and entanglement entropy in the corresponding quench \cite{Leichenauer:2016rxw}, which is hard to interpret physically. This additional restriction together with the NEC across the shell, $e_\text{dust}(z)\geq 0,\, p_\text{dust}(z)\geq 0$ \eqref{NECShell}, then implies:
\es{NECShell2}{
\delta a&\equiv a(z)-1=O(z^d)\,,\\
\delta b &\equiv b(z)-f(z)=O(z^d)\,.
}
We will use these below. We remark that the first relation implies that the source function $s(z)$ defined in \eqref{NECBH} is regular near the AdS boundary. The regularity of $b(z)$ at the boundary demands the other source function $t(z)$ to also be regular. However, in other contexts  it is necessary to consider a singular $s(z)$: the simplest example may be a hairy Reissner-Nordstrom black brane in Einstein-Maxwell gravity coupled minimally to a complex charged scalar.

Now we turn to the entanglement entropy, which we compute for entangling surface $\Sig$ of arbitrary shape. We compute the area of the codimension-two HRT surface, defined by the  general embedding $v=v(z,y^i)$ and ${\bf x}=\vec{x}(z,y^i)$, where $y^i$ are coordinates that parametrize $\Sig$ in the boundary theory and that we extend into the bulk. In this parametrization, the area functional is:
\begin{equation}
S(t) = \frac{1}{4G_N}\int_0^{z_t}dz \,d^{d-2}y\  \frac{1}{z^{d-1}}\sqrt{\det h}\,
\end{equation}
where $\frac{1}{z^2}\,h_{\alpha\beta}$ is the induced metric on the HRT surface, and $z_t$ is the point of deepest penetration of the surface into the bulk.

The early time behavior is characterized by the region of spacetime in which the null shell is still close to the boundary and most of the HRT surface lives in the spacetime \eqref{DomainWall} behind the shell. We thus regard the Vaidya geometry as a perturbation of  the spacetime \eqref{DomainWall}, and can get the early time expansion of the entropy by taking the static HRT surface that computes the entropy before the quench and lives entirely in the region  \eqref{DomainWall}, and plugging it into the deformed action functional \cite{Liu1,Liu2}. Since the HRT surface is extremal, its change due to the chance in geometry only contributes to the area at the next order. Let us denote by $z_c$ the point where the HRT surface crosses the null shell. While in general $z_c$ depends on $y^i$, at early times $t= z_c + O(z_c^2)$. We can then expand the area functional as
\begin{equation}
\hat{S}(t) = \int_0^{z_c} dz \,d^{d-2}y\ \left(\frac{\delta\mathcal{L}}{\delta a}\delta a + \frac{\delta \mathcal{L}}{\delta b}\delta b\right)\,,
\end{equation}
where $\mathcal{L} \equiv \frac{1}{z^{d-1}}\sqrt{\det h}$. We subtracted $S(0)$ to get $\hat{S}(t)$, see below \eqref{tSInts2}.   Expanding the determinant $\sqrt{\det h}$, we arrive at
\es{StVaidya}{
\hat{S}(t) = \frac{1}{8G_N}A_{\Sigma}\int_0^{z_c}\frac{dz}{z^{d-1}}\left(-\delta a - 2\delta b \right)\,,
}
where $\delta a,\,\delta b $ were defined in \eqref{NECShell2}. There we showed that both terms are $O(z^d)$, hence  the integral is convergent. The NEC implies that $b,f>0$, hence from \eqref{NECShell} also that  $\delta b\geq 0$. We can then bound the entropy from above by dropping the $- 2\delta b$ term from \eqref{StVaidya}, and plug in the leading small $z$ behavior of $-\de a$ to get:
\begin{equation}
\hat{S}(t) \leq \frac{1}{8G_N}A_{\Sigma}\int_0^{z_c}\frac{dz}{z^{d-1}}\left(a_d z^d \right)+\cdots\,.
\end{equation}
Applying the holographic dictionary to relate $a_d$ with $(e+p)$ \eqref{epHoloApp2}, and using that to leading order $z_c=t$, we get:
\begin{equation}
\hat{S}(t) \leq \frac{\pi(e+p)A_{\Sigma}t^2}{d}+\cdots\,.
\end{equation}
Thus
\begin{equation}
s_2^{\text{(holo)}} \leq \frac{\pi(e+p)}{d}\,.
\end{equation}
As emphasized in the main next, this bound is stronger by a factor of $d$ than the general field theory bound derived  from the QNEC. We conclude that the Vaidya quench never saturates the QNEC bound \eqref{s2Bound}.




\end{document}